\documentstyle[emulateapj,psfig,apjfonts]{article}
\def\gs{\mathrel{\raise0.35ex\hbox{$\scriptstyle >$}\kern-0.6em
\lower0.40ex\hbox{{$\scriptstyle \sim$}}}}
\def\ls{\mathrel{\raise0.35ex\hbox{$\scriptstyle <$}\kern-0.6em
\lower0.40ex\hbox{{$\scriptstyle \sim$}}}}

\submitted{Received 2001 -- --; Accepted: 2001 -- --}

\lefthead{Smith et al.}
\righthead{An {\it ISO}-detected ERO at $z=1.05$}

\begin{document}

\title{Near-infrared Spectroscopy and {\it Hubble Space Telescope} 
Imaging \\ of a  Dusty Starburst ERO\footnotemark}

\author{
Graham P.\ Smith,$\!$\altaffilmark{2,3}
Tommaso Treu,$\!$\altaffilmark{4}
Richard Ellis,$\!$\altaffilmark{4}
Ian Smail,\altaffilmark{2}
J.-P.\ Kneib\altaffilmark{5}
\& B.\ L.\ Frye.$\!$\altaffilmark{6}
}

\setcounter{footnote}{1}

\footnotetext{Based on observations with the NASA/ESA {\it Hubble Space
Telescope} obtained at the Space Telescope Science Institute, which is
operated by the Association of Universities for Research in Astronomy
Inc., under NASA contract NAS 5-26555.}
\altaffiltext{2}{Department of Physics, University of Durham,
South Road, Durham DH1 3LE, UK}
\altaffiltext{3}{E-mail: graham.smith@durham.ac.uk}
\altaffiltext{4}{Astronomy 105-24, Caltech, Pasadena CA 91125, USA}
\altaffiltext{5}{Observatoire Midi-Pyr\'en\'ees, CNRS-UMR5572,
14 Avenue E.\,Belin, 31400 Toulouse, France}
\altaffiltext{6}{Lawrence Berkeley National Laboratory, Department of 
Physics, Mailstop 50-232, Berkeley, CA 94720}

\setcounter{footnote}{6}

\begin{abstract}
We present near-infrared spectroscopy  and {\it Hubble Space Telescope
(HST)}  imaging of  ERO\,J164023+4644, an  Extremely Red  Object (ERO)
with $(R-K)=5.9$  at $z=1.05$ that has  been detected by  {\it ISO} at
15$\mu$m.   ERO\,J164023  appears  to   be  a  disk  galaxy,  with  an
optical/infrared  spectral  energy distribution  which  appears to  be
strongly  reddened  by dust  ($L_{\sc  fir}/L_{\sc b}\ls200$;  $A_{\sc
v}\sim5$).  The combination of the  narrow width of the emission lines
in our  spectra ($\sim300$kms$^{-1}$)  and the relatively  high [N{\sc
ii}]/H$\alpha$  line  ratio  indicate  that this  is  a  ``composite''
starburst-Seyfert galaxy.  Assuming  that star formation dominates the
energy  output, we constrain  the star  formation rate  to lie  in the
broad range $\sim10$--700\,M$_{\odot}$yr$^{-1}$ from a variety of star
formation  indicators.  We  compare ERO\,J164023  with the  only other
spectroscopically   identified  dusty   EROs:   HR10  ($z=1.44$)   and
ISO\,J1324$-$2016  ($z=1.50$).   ERO\,J164023  and HR10  have  similar
disk-like  morphologies in  the  rest  frame UV,  and  both exhibit  a
variation  in  the  apparent   dust  obscuration  depending  upon  the
diagnostic used which suggests that  there is a complex spatial mix of
stellar  populations and  dust in  these galaxies.   In  contrast, the
compact  morphology  and   spectral  properties  of  ISO\,J1324$-$2016
indicate that it is a  dusty quasar.  Overall, our results demonstrate
that the population of dusty galaxies identified using photometric ERO
criteria  includes  systems  ranging  from  pure  starbursts,  through
transition systems such as  ERO\,J164023 to dusty quasars.  We suggest
that the classification of  EROs into these sub-classes, necessary for
the detailed modelling of  the population, cannot be reliably achieved
from    optical/near-infrared   photometry   and    instead   requires
mid/far-infrared   or  sub-millimetre  photometry   and  near-infrared
spectroscopy.   The  advent  of efficient  multi-object  spectrographs
working in  the near-infrared as well  as the imminent  launch of {\it
SIRTF}  therefore promise  the opportunity  of rapid  progress  in our
understanding of the elusive ERO population.
\end{abstract}


\section{Introduction}

Our understanding of  both the star formation history  of the Universe
and emission from  active galactic nuclei (AGN) was  challenged in the
late-1990s by  mid-infrared (MIR) (e.g.\ Rowan-Robinson  et al.\ 1997;
Flores et al.\ 1999) and  sub-millimetre (sub-mm) (Smail et al.\ 1997,
2001;  Eales  et al.\  1999;  Dunlop  2001)  observations of  luminous
far-infrared  (FIR)  galaxies  at  high  redshift.  If,  as  has  been
suggested,  the FIR  emission  from these  galaxies  arises from  star
formation then these observations suggest that around half of all star
formation may have occurred in  highly obscured systems (Blain et al.\
1999;  Elbaz et al.\  1999).  Tracing  the evolution  of dust-obscured
star-formation  and AGN-activity  as  a function  of  cosmic epoch  is
therefore essential to our understanding of galaxy evolution.

The identification of the optical counterparts of these sub-mm and MIR
sources is not straightforward due to the optical faintness of many of
the candidates  and the large  positional uncertainties in  the longer
wavelength   observations   ($\sim3$--4$''$).    Nevertheless,   where
counterparts  are found, many  of these  appear to  be red  in optical
passbands, with some being extremely red objects (EROs) with $(R-K)>6$
(Smail et  al.\ 1999; Gear et  al.\ 2000).  Whilst it  is possible for
such colors to be produced  by evolved stellar populations at $z\gs1$,
the intense FIR emission of  these galaxies suggests that dust is also
likely to be responsible for their extreme colors.

Recent imaging observations suggest that  up to 20--50 per cent of all
EROs with  $(R-K)\ge5.3$ and the majority of  those with $(R-K)\ge6.0$
may be dusty starburst galaxies (Smail et al.\ 1999; Treu \& Stiavelli
1999;  Moriondo  et al.\  2000;  Smith  et  al.\ 2001).   However,  to
understand  more   about  the   nature  of  these   galaxies  requires
spectroscopic  study.  Such  observations  are extremely  challenging,
even with 10-m class telescopes,  due to the optical faintness of EROs
($R\gs$24).    Consequently,   only   two   dusty   EROs   have   been
spectroscopically  identified to  date:  HR10, a  starburst galaxy  at
$z=1.44$ (Dey et  al.\ 1999 - hereafter D99)  and ISO\,J1324$-$2016, a
dusty AGN at $z=1.50$ (Pierre et al.\ 2001).

To  drive forward  this  field  Smith et  al.\  (2001, hereafter  S01)
recently exploited the magnifying  power of ten massive galaxy cluster
lenses  to boost the  sensitivity of  follow-up spectroscopy  of EROs.
They  combined  ground-based  NIR  observations with  deep  {\it  HST}
optical  imaging to construct  a sample  of 60  gravitationally lensed
EROs.  This paper  presents the detailed follow-up of  one source from
this   sample:    ERO\,J164023+4644   (hereafter   referred    to   as
ERO\,J164023).   This galaxy  was previously  associated with  an {\it
ISO}  15$\mu$m source  by Barvainis  et al.\  (1999),  who tentatively
identified a single optical emission  line as [O{\sc ii}] at $z=1.05$,
although they did not realise that the galaxy is an ERO.

We describe  our observations in \S2  and present our  results in \S3.
We  compare ERO\,J164023 with  HR10 and  ISO\,J1324$-$2016 in  \S4 and
summarise     our      conclusions     in     \S5.       We     assume
$H_0=50$kms$^{-1}$Mpc$^{-1}$, $\Omega_0$=1 and $\Lambda_0=0$ and adopt
a lens amplification of 1.4 for ERO\,J164023 (S01).

\medskip
%
%
\centerline{\psfig{file=fig1a.ps,width=1.65in,angle=-90}
\hspace{2mm}
\psfig{file=fig1b.ps,width=1.65in,angle=-90}}
\smallskip
\noindent{\scriptsize \addtolength{\baselineskip}{-3pt}
{\sc Fig.~1.} ---
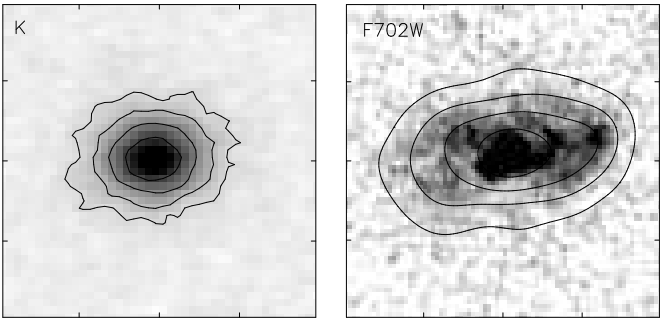
The  UKIRT $K$-band  (left)  and  {\it HST}  F702W  (right) images  of
ERO\,J164023.  The  resolution of these two images  is $\sim0.5''$ and
$\sim0.1''$  respectively.  The  contours on  the left-hand  panel are
from the  raw $K$-band  image that also  appears as the  gray-scale in
this  panel and  the contours  in the  right-hand panel  are  from the
seeing-matched  (i.e.\ $\sim0.5''$ resolution)  {\it HST}  frame.  The
tick marks are $1''$ apart; North is up and East is Left.

}
\medskip

\section{Observations}

\subsection{Imaging}

The  $K$-band data  used to  identify ERO\,J164023  was obtained  in a
6.5-ks  exposure  using the  UFTI  imager  on  the 3.8-m  UK  Infrared
Telescope\footnote{The United  Kingdom Infrared Telescope  is operated
by the  Joint Astronomy Centre on  behalf of the  Particle Physics and
Astronomy Research Council} (UKIRT) in  0.5$''$ seeing on 2000 April 5
(S01).   ERO\,J164023 lies  at 16:40:23.05  +46:44:02.3 (J2000)  and a
$K$-band image of this galaxy is shown in Fig.~1.

In addition to  the $K$-band imaging, we obtained  a $J$-band image of
ERO\,J164023  with  INGRID on  the  4.2-m  William Herschel  Telescope
(WHT)\footnote{The William Herschel Telescope is operated by the Isaac
Newton  Group on  behalf  of Particle  Physics  \& Astronomy  Research
Council}  on  2001  May  6.   This  observation  totalled  2.2\,ks  in
$\sim0.8''$  seeing,  and was  reduced  in  a  similar manner  to  the
$K$-band data.   We also exploit  $H$-band observations of  this field
from Gray et al.\ (2000) using CIRSI on the WHT.

Morphological information on  the ERO comes from {\it  HST} imaging of
the cluster.  The field containing ERO\,J164023, A\,2219, was observed
with  {\it  WFPC2}  onboard  {\it  HST} for  six  exposures  totalling
14.4\,ks  through the  F702W filter  (S01).  We  show the  {\it WFPC2}
image of this  galaxy in Fig.~1.  To obtain  optical photometry of the
field we  have also analysed  archival $UBVI$-band imaging  of A\,2219
taken  with COSMIC  on  the Hale  5-m\footnote{The  Hale Telescope  at
Palomar Observatory is owned  and operated by the California Institute
of Technology} and LRIS  on Keck\footnote{The W.\,M.\ Keck Observatory
which  is operated as  a scientific  partnership between  Caltech, the
University of California and NASA} (Smail et al.\ 1995, 1998).

Mid-infrared observations  of the galaxy  come from Barvainis  et al.\
(1999) who  observed A\,2219  at $15\mu$m  with {\it  ISOCAM} on-board
ESA's  {\it Infrared  Space  Observatory (ISO)}.   They detected  five
sources,  one   of  which   (A2219\#5)  lies  within   $\sim0.8''$  of
ERO\,J164023.

At longer  wavelengths, A\,2219 was  also observed by Chapman  et al.\
(2000) using SCUBA  at $850\mu$m  (Holland et al.\  1999) on  the 15-m
James  Clerk   Maxwell  Telescope\footnote{The  James   Clerk  Maxwell
Telescope is operated by the JAC  on behalf of the Particle Physics \&
Astronomy  Council,   the  Netherlands  Organisation   for  Scientific
Research  and   the  National  Research  Council   of  Canada}.   They
tentatively   identified   one   of   the  sources   in   this   field
(SMM\,J16404+4644)  with  A2219\#5  from  Barvainis  et  al.\  (1999).
However, as  the 850$\mu$m source is $\gs6''$  away from ERO\,J164023,
we suggest that this  identification is probably incorrect and instead
we adopt Chapman et al.'s 3-$\sigma$ detection as a conservative upper
limit on the 850$\mu$m flux from ERO\,J164023.

Finally, we obtain  3-$\sigma$ detection limits in the  radio from the
28.5\,GHz,  4.9\,GHz, 1.4\,GHz  maps of  Cooray et  al.\  (1998), Edge
(private communication) and Owen (private communication).  We list our
optical, infrared, sub-millimeter and radio photometry in Table~1.

%
%
\centerline{\sc Table 1}
\centerline{\sc Photometry of ERO\,J164023}
\vspace{0.1cm}
{\scriptsize
\begin{center}
\begin{tabular}{cccl}
\hline\hline
\noalign{\smallskip}
{Observed} &  {Flux Density$^{a,c}$} & {Magnitude$^{b,c}$} & {Reference} \cr
{Band} & ($\mu$Jy) & & \cr
\noalign{\smallskip}
\hline
\noalign{\smallskip}
$U$ & $<0.10$ & $>25.6$ & Smail et al.\ (1998) \cr
$B$ & $0.13\pm0.02$ & $26.2\pm0.1$ & Smail et al.\ (1998) \cr
$V$ & $0.19\pm0.03$ & $25.7\pm0.2$ & Smail et al.\ (1995) \cr
$R$ & $1.32\pm0.05$ & $23.54\pm0.04$ & S01 \cr
$I$ & $3.00\pm0.24$ & $22.15\pm0.08$ & S01 \cr
$J$ & $20.4\pm0.6$  & $19.72\pm0.06$ & This paper \cr
$H$ & $21.3\pm0.7$  & $19.20\pm0.09$ & Gray et al.\ (2000) \cr
$K$ &  $63.0\pm0.6$ & $17.64\pm0.01$ & S01 \cr
15$\mu$m & $530\pm110$ & ... & Barvainis et al.\ (1999) \cr
850$\mu$m & $<6\times10^3$ & ... & Chapman et al.\ (2000) \cr
28.5\,GHz & $<780$ & ... & Cooray et al.\ (1998) \cr
4.9\,GHz & $<300$ & ... & Edge (private communication) \cr
1.4\,GHz & $<100$ & ... & Owen (private communication) \cr
\hline
\end{tabular}
\end{center}
\addtolength{\baselineskip}{-3pt}
$^a$ Errors quoted are 1-$\sigma$. Limits quoted are 3-$\sigma$ detection 
limits.\newline
$^b$ All magnitudes are measured in a $2''$ diameter aperture.\newline
$^c$ Flux density measurements and magnitudes in this table have not been 
corrected for gravitational amplification by the foreground cluster lens, 
A\,2219.

}
\medskip

\subsection{Spectroscopy}

A  NIR spectrum  of ERO\,J164023  was  obtained in  the $J$-band  with
NIRSPEC  on Keck-II  on 2001  April  9 in  photometric conditions  and
$0.6''$ seeing.  Three exposures of 600\,s each were obtained, nodding
along the $0.76''\times42''$ slit by $5''$ between each exposure.  The
resolution of these observations was $10$ \AA\ FWHM.

\medskip
%
%
\centerline{\psfig{file=fig2.ps,width=3.5in,angle=-90}}
\smallskip
\noindent{\scriptsize \addtolength{\baselineskip}{-3pt}
{\sc Fig.~2.} --- 
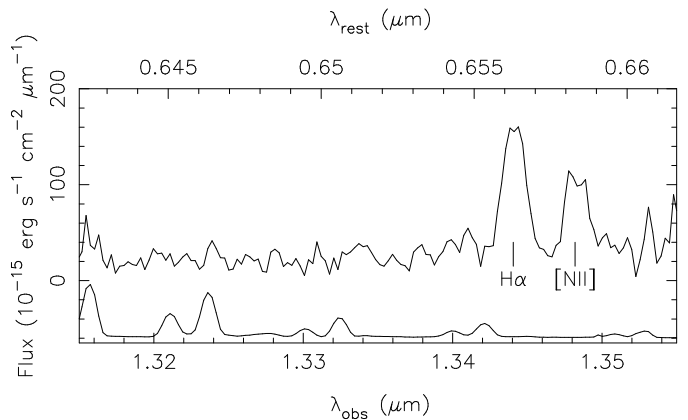
The  near-infrared spectrum  of ERO\,J164023.  We identify  the strong
emission    lines    as    H$\alpha$    $\lambda$6563    and    [N{\sc
ii}]$\lambda$6583, giving a redshift of $z=1.0480\pm0.0005$.

}
\medskip

Wavelength  calibration  was  achieved   from  the  sky  lines,  using
observations of  a bright star to correct  for geometrical distortion.
An  average sky  spectrum was  created  by scaling  and combining  the
individual  spectra and  this was  then subtracted  from  each science
observation   before   co-adding   the   spectra  and   extracting   a
one-dimensional  spectrum.   Flux  calibration  was  achieved  through
observations of UKIRT standard stars,  which were also used to correct
for  telluric absorption.  No attempt  was  made to  correct for  slit
losses.

\medskip
%
%
\centerline{\psfig{file=fig3.ps,width=3.5in,angle=-90}}
\smallskip
\noindent{\scriptsize \addtolength{\baselineskip}{-3pt}
{\sc Fig.~3.} --- 
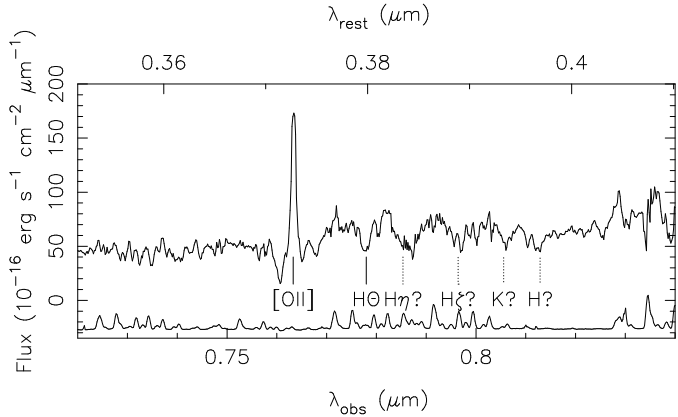
The optical spectrum of  ERO\,J164023. We identify the strong emission
line   as  [O{\sc   ii}]$\lambda$3727  which   is  confirmed   by  the
identification  of the  H$\alpha$ and  [N{\sc  ii}] lines  in the  NIR
spectrum  (Fig.~2).   The H${\rm\theta}$  absorption  feature is  also
tentatively detected, although other  members of the Balmer series and
the Ca\,H  and Ca\,K  features are not  visible.  The sky  spectrum is
also showed in arbitrary units below the science spectrum.

}
\medskip

The flux-calibrated spectrum (Fig.~2) reveals two strong emission lines
at  1.3441 and $1.3483\mu$m, which we identify as H$\alpha$, and [N{\sc
ii}]$\lambda$6583 respectively. This places the galaxy at a redshift of
$z=1.0480\pm0.0005$.

Barvainis  et al.\  (1999)  discuss an  optical  spectrum of  A2219\#5
(\S2.1) obtained by  Frye \& Broadhurst using LRIS  on Keck-I which is
shown in Fig.~3.  This shows  a strong emission line at 7634\AA\ which
Barvainis et  al.\ tentatively identified  as [O{\sc ii}]$\lambda$3727
at a  redshift of  $z=1.048$, as confirmed  by our NIR  spectrum.  The
optical spectrum also appears to contain the H${\rm\theta}$ absorption
feature,  although  other  members   of  the  Balmer  series  and  the
Ca\,H\,\&\,K  lines are  not  convincingly detected.   We present  the
emission line measurements in Table~2.
 
%
%
%
\centerline{\sc Table 2}
\centerline{\sc Emission Line Measurements of ERO\,J164023}
\vspace{0.1cm}
{\scriptsize
\begin{center}
\begin{tabular}{lcccc}
\hline\hline
\noalign{\smallskip}
{Line} & $\lambda_{obs}$ & {Flux $^a$} & {FWHM$^{a,b}$} & $W_{\lambda, rest}^a$ \cr
& ($\mu$m) & ($10^{-17}$erg\,s$^{-1}$cm$^{-2}$) & (km\,s$^{-1}$) & (\AA) \cr
\noalign{\smallskip}\hline\noalign{\smallskip}
[O{\sc ii}] & 0.7632 & $2\pm1$ & $<320$ & $30\pm5$ \cr  
H$\alpha$ & 1.3441 & $27\pm2$ & $310\pm50$ & $60\pm15$ \cr
[N{\sc ii}] & 1.3483 & $18\pm2$ & $330\pm50$ & $50\pm20$ \cr
\noalign{\smallskip}
\hline
\noalign{\smallskip}
\end{tabular}
\end{center}
\addtolength{\baselineskip}{-3pt}
$^a$ Fluxes, equivalent widths, and FWHMs are measured by fitting a 
Gaussian using {\sc splot} in {\sc iraf} and are not corrected for
gravitational amplification.
\newline $^b$ FWHM measurements are quoted after correcting for the 
instrumental resolution of the spectrograph.

}


\section{Results and Discussion}

\subsection{Morphology}

The   {\it   HST}  frame   ($\sim0.1''$   resolution)  suggests   that
ERO\,J164023 has a disk morphology in the rest frame UV (Fig.~1), with
no obvious signs of  strong dynamical disturbance.  The disk component
is  sufficiently  extended,  $\sim  2$--3$''$ (10--15\,kpc),  that  it
remains  visible  when  the  {\it   HST}  frame  is  degraded  to  the
ground-based  resolution ($0.5''$).  In  contrast, at  this resolution
the  $K$-band emission  shows  less evidence  of  a disk  and is  more
concentrated.   The $(R-K)\sim6$  color of  ERO\,J164023  (Table~1) is
dominated by its central regions,  with the outer regions displaying a
more ``modest'' optical/NIR colour of $(R-K)\sim5.3$.

\medskip
%
%
\centerline{\psfig{file=fig4.ps,width=3.2in,angle=-90}}
\smallskip
\noindent{\scriptsize \addtolength{\baselineskip}{-3pt}
{\sc Fig.~4.} ---
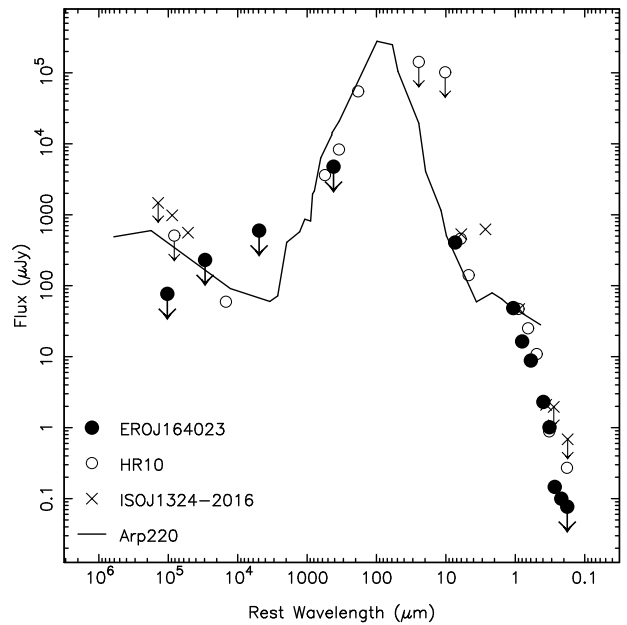
The rest frame spectral energy distribution (SED) of ERO\,J164023 from
optical to radio wave-bands.  For comparison, we show the SEDs of HR10
(D99; Elbaz et al.\ 2001), ISO\,J1324$-$2016 (Pierre et al.\ 2001) and
Arp220 (Hughes et al.\ 1998) The flux densities of ERO\,J164023 have all
been corrected for gravitational amplification (\S1).  The SEDs of all
four galaxies have been normalised to the observed $K$-band flux
of ERO\,J164023.

}
\medskip

\subsection{Spectral Energy Distribution}

We  plot  the  rest   frame  spectral  energy  distribution  (SED)  of
ERO\,J164023  in Fig.~4.  The  SED of  ERO\,J164023  rises steeply  at
optical/MIR  wavelengths   and,  although  less   well-constrained  at
sub-mm/radio  wavelengths,  the flux  density  probably  peaks in  the
FIR/sub-mm.

The  extreme  redness  of  the  SED  at  optical/NIR/MIR  wavelengths,
suggests  that the  spectrum of  ERO\,J164023 is  heavily  reddened by
dust.   We quantify  this  in  two ways,  first  using the  rest-frame
FIR-to-blue  luminosity ratio  ($L_{\sc fir}/L_B$).   We  estimate the
(unlensed) FIR  luminosity of  ERO\,J164023, using the  850$\mu$m flux
limit (Table~1) to  scale the FIR luminosity of  HR10 (D99), obtaining
$L_{\sc fir}\ls4\times10^{12}L_{\odot}$, and we measure the (unlensed)
rest-frame blue  luminosity to be  $L_B\sim2\times10^{10}L_{\odot}$ by
interpolating between the observed  $I$- and $J$- bands (Table~1).  We
therefore  estimate  $L_{\sc  fir}/L_B\ls200$.   Second, we  use  {\sc
hyper-z}  (Bolzonella  et  al.\   2000)  to  investigate  the  SED  of
ERO\,J164023  by fitting  template  star-forming model  spectra for  a
range of  star formation ages  and dust extinction to  the photometric
data presented  in Table~1.  Assuming  a starburst age  of $<0.2$\,Gyr
and  adopting  the   spectroscopic  redshift  ($z=1.048$)  produces  a
solution of $A_{\sc v}\sim5$,  confirming that ERO\,J164023 is heavily
dust-obscured.

\subsection{Starburst or AGN?}

The  spectra of  ERO\,J164023  (Fig.~2) contains  narrow [O{\sc  ii}],
H$\alpha$  and  [N{\sc  ii}]  emission  lines  (Table~2)  and  a  weak
H$\theta$  absorption feature.  These spectral  features are  all more
typical  of star  forming galaxies  than AGN  (e.g.\ Liu  \& Kennicutt
1995).
  
However,  the   [N{\sc  ii}]  to  H$\alpha$   flux  ratio,  log([N{\sc
ii}]/H$\alpha)\sim-0.2$  (Table~2), is higher  than the  typical value
($-0.5$) seen for star forming  galaxies and suggests that the ERO may
be  a  Seyfert 2  (Veilleux  \& Osterbrock  1987)  or  more likely,  a
composite starburst-Seyfert galaxy (Hill et al.\ 1999).  Nevertheless,
the  nuclear activity in  ERO\,J164023 may  be quite  weak, as  an AGN
contribution to the line emission of  as little as 10 per cent appears
to be  sufficient to  account for the  [N{\sc ii}] to  H$\alpha$ ratio
(Hill et al.\ 2001).  We  also note that ERO\,J164023 was not detected
when A\,2219  was observed  at X-ray wavelengths  by {\it  ROSAT HRI},
implying that any  AGN is not X-ray bright.   The 3-$\sigma$ detection
limit    from   these    observations    is   $<5.7\times10^{-4}\mu$Jy
[0.1--2.4\,keV] (Edge, private communication).

Assuming that star formation  dominates the line emission, we estimate
the   star   formation   rate   (SFR)   of   ERO\,J164023   from   its
lensing-corrected         H$\alpha$         luminosity:        $L_{\rm
H\alpha}=3.3\times10^8\,L_{\odot}$,       obtaining       SFR$_{\rm
H\alpha}\sim10$\,M$_{\odot}$yr$^{-1}$  for a  Salpeter  IMF (Kennicutt
1998).  As  we have made no  corrections for dust  extinction, this is
probably a  lower limit.  Indeed,  a suppression of the  observed line
fluxes due  to dust is also suggested  by the very low  [O{\sc ii}] to
H$\alpha$ flux ratio of  ERO\,J164023 ($0.07\pm 0.03$, Table~2), which
is far  lower than the value  seen in nearby  spirals, $0.43\pm 0.27$,
and more  typical of local  dusty starburst galaxies (Poggianti  \& Wu
2000).

The constraints on the luminosity of the ERO at longer wavelengths can
also be  used to place an  upper limit on the  probable star formation
rate.   Based  on  the  lensing-corrected far-infrared  luminosity  of
$L_{\sc   fir}\ls4\times10^{12}L_{\odot}$,   we   estimate   SFR$_{\sc
fir}\ls700$\,M$_{\odot}$yr$^{-1}$  (Kennicutt 1998).  A  similar upper
limit  to  the  SFR is  obtained  based  on  the 1.4\,GHz  flux  limit
(Table~1).  We therefore can only constrain the SFR of ERO\,J164023 to
lie   within  a   broad   range:  $\sim10$--700\,M$_{\odot}$yr$^{-1}$,
although all  of the  estimates suggest that  this is a  strongly star
forming  galaxy.  The  factor of  70  difference between  the FIR  and
H$\alpha$  estimates of  the  SFR  is consistent  with  that found  by
Poggianti \&  Wu (2000) for local starburst  galaxies, indicating that
the SFR could feasibly be as high as $\sim$700\,M$_{\odot}$yr$^{-1}$.

\section{The dusty ERO population}

We  now compare  the properties  of ERO\,J164023  with the  only other
spectroscopically-confirmed dusty EROs: the $z=1.44$ starburst galaxy,
HR10 (D99), and the dusty quasar ISO\,J1324$-$2016 at $z=1.50$ (Pierre
et al.\ 2001).

We start by classifying the EROs using the $(I-K)$--$(J-K)$ diagnostic
diagram  suggested  by Pozzetti  \&  Mannucci (2000).   Unfortunately,
there is  no published $J$-band photometry  for ISO\,J1324$-$2016, but
using the $(I-K)$ and  $(J-K)$ colors of ERO\,J164023 ($4.5\pm0.1$ and
$2.1\pm0.1$; S01) and HR10  ($5.8\pm0.1$ and $2.6\pm0.1$; D99) we find
that  these  galaxies both  lie  just on  the  starburst  side of  the
proposed  dividing line  between  evolved and  starburst  EROs on  the
$(I-K)$--$(J-K)$   plane.     However,   due   to    the   photometric
uncertainties, neither  ERO can be robustly classified  as a starburst
on the basis of these three photometric bands alone.

To compare  the three EROs across  a broader wavelength  range we plot
the SEDs of HR10 and  ISO\,J1324$-$2016 in Fig.~4.  Dealing first with
HR10   --  the   spectral   shapes  of   ERO\,J164023   and  HR10   at
optical/NIR/MIR wavelengths are very similar and both resemble the SED
of  Arp\,220, implying that  both galaxies  are highly  obscured.  D99
estimate $L_{\sc fir}/L_B\sim300$ for  HR10, suggesting that it may be
more heavily obscured than ERO\,J164023.  On the other hand, the crude
estimates of dust obscuration  from the $V$-band extinction and [O{\sc
ii}]/H$\alpha$ line  ratio suggest  that ERO\,J164023 and  HR10 suffer
comparable  degrees of  dust  obscuration.  The  scatter  seen in  the
estimates of  the relative extinction  in these two  systems depending
upon the diagnostic used supports the idea that the dust has a complex
spatial distribution and produces different degrees of obscuration for
the emission from the  various stellar populations within the galaxies
(e.g.\ Poggianti \& Wu 2000).

There  is some morphological  support for  differences in  the spatial
distribution of visible stars and  dust in both HR10 and ERO\,J164023.
{\it HST} imaging of HR10 in the F814W filter is discussed by D99, the
higher  redshift of  this  ERO means  that  these observations  sample
similar rest frame  UV wavelengths to the F702W  image of ERO\,J164023
($\sim3500$\AA).    HR10   displays   an   ``S''--shaped   morphology,
suggesting that it is a disk  galaxy, although D99 propose that it may
be  dynamically disturbed.   In contrast,  the $K$-band  morphology of
HR10 is  more symmetrical, suggesting that, like  ERO\,J164023, it may
either  harbor  a bulge  of  older stars,  or  more  likely a  central
starburst that is more heavily dust-reddened than the outskirts of the
galaxies.

D99 propose that HR10 is a dusty starburst-powered galaxy based on the
lack  of signatures  of nuclear  activity  in both  its emission  line
widths   ($\sim600$\,km\,s$^{-1}$)  and   its  emission   line  ratio:
log([N{\sc ii}]/H$\alpha)\sim-0.4$.  Applying  the same conversions as
used  in  \S3.3,  we estimate  the  SFR  in  HR10  lies in  the  range
$\sim40$--1200\,M$_{\odot}$yr$^{-1}$,  where  the  lower  bound  again
comes from the observed H$\alpha$  flux (uncorrected for dust) and the
upper bound is based on the  FIR luminosity of the galaxy.  This range
suggests that  HR10 may  be forming stars  at a rate  $\sim$2--3 times
higher than ERO\,J164023.

Turning now to the  comparison of ERO\,J164023 with ISO\,J1324$-$2016,
we see that in contrast to the broad similarities between ERO\,J164023
and    HR10,   the    galaxy   shares    few    characteristics   with
ISO\,J1324$-$2016.  The lack of {\it HST} imaging of ISO\,J1324$-$2016
precludes detailed analysis of  its morphology, however Pierre et al.\
(2001) state that this source appears point-like in $\sim0.5''$ seeing
in the $K$-band,  unlike ERO\,J164023 (and HR10) which  is extended in
similar  seeing at  these wavelengths  (Fig.~1).  The  optical  SED of
ISO\,J1324$-$2016  also  shows  the  precipitous  decline  in  the  UV
characteristic of  reddening by dust  (Pierre et al.\  (2001) estimate
$A_V\sim 4$--7 from the  Balmer decrement in ISO\,J132402916), however
the behaviour in the  mid-infrared is very different from ERO\,J164023
(and  HR10)  with excess  emission  at  6.75\,$\mu$m,  which has  been
interpreted   as  evidence   for  a   hot  component,   probably  from
circumnuclear dust around an AGN.

The identification of ISO\,J1324$-$2016 as a dusty quasar is confirmed
by a strong,  broad H$\alpha$ emission line ($\sim3000$\,km\,s$^{-1}$)
and      relatively      strong      radio      emission      (P$_{\rm
1.4\,GHz}\sim2\times10^{25}$\,W\,Hz$^{-1}$).   This contrasts markedly
with the  narrow line widths seen  in both ERO\,J164023  and HR10.  In
addition, neither ERO\,J164023, nor HR10, have so far been detected at
1.4\,GHz, although  both appear to be  at least an  order of magnitude
less  luminous  than  ISO\,J1324$-$2016  in this  band,  with  P$_{\rm
1.4\,GHz}\ls5\times10^{23}$   and   $\ls4\times10^{24}$   W\,Hz$^{-1}$
respectively.

In  summary: the  properties of  ERO\,J164023 suggest  it  shares many
characteristics with the ultraluminous  starburst HR10, although it is
less extreme  and also shows  evidence of nuclear  activity.  However,
the AGN component in ERO\,J164023 is dominated in most observations by
the  star formation  in  this  galaxy and  hence  this system  differs
markedly from the dusty quasar, ISO\,J1324$-$2016.

\section{Conclusions}

We   have   obtained   a   secure  spectroscopic   identification   of
ERO\,J164023, a dusty starburst-Seyfert ERO at a redshift of $z=1.05$.
This  brings the  total number  of spectroscopically  identified dusty
EROs  to  three and  adds  a  composite  AGN/starburst system  to  the
apparently   pure   starburst   (HR10)   and   AGN-dominated   systems
(ISO\,J1324$-$2016) previously identified.

We complement  this spectroscopy with  deep {\it HST}  optical imaging
and  infrared,  sub-mm  and  radio  data, enabling  us  to  study  the
morphology and  SED of this  unusual galaxy.  The main  conclusions of
our work are as follows:

(i)  ERO\,J164023  is a  disk  galaxy,  the  central region  of  which
dominates its optical/NIR color of $(R-K)\sim6$.  The steep optical/IR
SED  is consistent  with this  galaxy being  heavily obscured  by dust
($L_{\sc fir}/L_B\ls200$; $A_{\sc v}\sim5$).

(ii) The  spectral line widths  are consistent with the  dust emission
being powered by hot  young stars.  However, the [N{\sc ii}]/H$\alpha$
line   strength  ratio   suggests   that  this   is  a   ``composite''
starburst-Seyfert galaxy.  Assuming star  formation to be the dominant
power  source,  we  constrain  the  SFR  to lie  in  the  broad  range
$\sim10$--700M$_{\odot}$yr$^{-1}$  which  is  a factor  of  $\sim$2--3
times lower than HR10.

(iii)  ERO\,J164023  and  HR10  have  strikingly  similar  rest  frame
optical/NIR/MIR  spectral   properties  and  both   exhibit  disk-like
morphologies.   The  dominant role  of  the  central  region of  these
galaxies  in  producing  their  extremely red  optical/NIR  colors  is
consistent with  them both containing an  obscured, central starburst.
Variation  of the measured  dust obscuration  suffered by  each galaxy
from  a number  of  diagnostics  suggests that  the  dust and  various
stellar populations  within dusty  starburst galaxies differ  in their
spatial distributions.

(iv) Despite the photometric similarities  of the three EROs, HR10 and
ERO\,J164023 both  differ from ISO\,J1324$-$2016 in  that the latter's
emission is primarily driven by an AGN, whereas star formation appears
to be the power source in HR10 and ERO\,J164023.

Overall, our work reveals that  samples of dusty EROs contain the full
range  of   power  sources:  dusty  starbursts,   AGNs  and  composite
starburst-AGN  systems.  The  broad  wavelength range  of our  imaging
(X-ray to radio) together with our optical and NIR spectroscopy allows
us to identify which observations are required to accurately segregate
active (i.e.\ dusty) and passive EROs and then to classify active EROs
into their different sub-classes.

Firstly,  whilst the $(I-K)$--$(J-K)$  color-color plane  (Pozzetti \&
Mannucci  2000) provides  a  rough classification  between active  and
passive  EROs,  photometric  uncertainties  appear  to  undermine  the
accuracy  of this  method (a  concern shared  by Pozzetti  \& Mannucci
2000).   Accurate  separation of  active  and  passive EROs  therefore
requires  MIR/FIR or submm  observations to  search for  the signature
dust  emission of  active  systems.  The  forthcoming  launch of  {\it
SIRTF}  should provide  the  opportunity for  rapid  progress in  this
respect.

However, NIR spectroscopy is also essential to further classify active
EROs into their respective sub-classes  on the basis of their detailed
spectral  properties.  Spectral analysis  of a  statistically reliable
sample of active  EROs is therefore crucial to  a better understanding
of the evolution of dust-obscured star-formation and AGN-activity as a
function  of  cosmic  epoch.    We  anticipate  that  the  forthcoming
generation of  NIR multi-object spectrographs will  play a significant
role in advancing our understanding of the elusive ERO population.

\section*{Acknowledgements}

We are especially grateful to Roser Pell\'o for her assistance with
the {\sc hyper-z} analysis and Karl Glazebrook for help with the Keck
observations.  We are also grateful to Scott Chapman, Alastair Edge,
Meghan Gray, Frazer Owen, for sharing their observational data with us.
Thanks also go to Harald Ebeling, Rob Ivison, Harald Kuntschner, Alice
Shapley and Chuck Steidel for a variety of invaluable discussions and
assistance.  GPS acknowledges a postgraduate studentship from PPARC. IRS
acknowledges support from the Royal Society and the Leverhulme Trust.
JPK acknowledges support from CNRS.  We also acknowledge support from
the UK--French ALLIANCE collaboration programme \#00161XM.


\begin{references}
\reference{} Barvainis R., Antonucci R., Helou G., 1999, AJ,
	118, 645
\reference{} Blain A.W., Smail I., Ivison R.J., Kneib J.-P., 1999, 
        MNRAS, 302, 632
\reference{} Bolzonella M., Miralles J.M., Pell\'o, 2000, A\&A, 363, 476
\reference{} Chapman S.C., Scott D., Borys C., Fahlman G.G., 2001, 
        MNRAS, submitted (astro-ph/0009067)
\reference{} Cooray A.R., Grego L., Holzapfel W.L., Joy M., Carlstrom 
        J.E., 1998, AJ, 115 1388
\reference{} Dey A., Graham J.R., Ivison R.J., Smail I.,
	Wright G.S., Liu M.C., 1999, ApJ, 519, 610 (D99)
\reference{} Dunlop J.S., 2001, in UMass/INAOE Conference on Deep 
        Millimeter Surveys, eds.\ Lowenthal J., Hughes D., (astro-ph/0011007)
\reference{} Eales S.A., Lilly S.J., Gear W.K., Dunne L., Bond J.R., 
        Hammer F., Le F\'evre O., Crampton D., 1999, ApJ, 515, 518
\reference{} Elbaz D., et al., 2001, A\&A, submitted
\reference{} Flores H., Hammer F., Thuan T.X., Ce\'sarsky C., Desert F.X., 
        Omont A., et al., 1999, ApJ, 517, 148
\reference{} Gear W.K., Lilly S.J., Stevens J.A., Clements D.L., Webb T.M., 
        Eales S.A., Dunne L., 2000, MNRAS, 316, 51
\reference{} Gray M.E., Ellis R.S., Refregier A., Be\'zecourt J., McMahon 
        R.G., Beckett M.G., Mackay C.D., Hoenig M.D., 2000, MNRAS, 318, 573
\reference{} Hill T.L., Heisler C.A., Sutherland R., Hunstead R.W., 1999, 
        AJ, 117, 111
\reference{} Hill T.L., Heisler C.A., Norris R.P., Reynolds J.E., Hunstead 
        R.W., 2001, 121, 128
\reference{} Holland W.S., Robson E.I., Gear W.K., Cunningham C.R., 
        Lightfoot J.F., Jenness T., et al., 1999, MNRAS, 303, 659
\reference{} Hughes D.H., Dunlop J.S., 1998, in Carilli C., et al., eds, 
        ASP Conf. Ser., Highly Redshifted radio lines.\ Soc.\ Pac.\ San 
        Fransisco.
\reference{} Kennicutt R.C., 1998, ARA\&A, 36, 189
\reference{} Liu C.T., Kennicutt R.C., 1995, ApJ, 450, 547
\reference{} Moriondo G., Cimatti A., Daddi E., 2000, A\& A, 364, 26
\reference{} Pierre M., Lidman C., Hunstead R., Alloin D., Casali M., 
        Cesarsky C., Chanial P., Duc P.-A., Fadda D., Flores H., Madden S., 
        Vigroux L., 2001, A\&A, submitted (astro-ph/0105075)
\reference{} Poggianti B.M., Wu H., 2000, ApJ, 529, 157
\reference{} Pozzetti L., Mannucci F., 2000, MNRAS, 317 L17
\reference{} Rowan-Robinson M., et al., 1997, MNRAS, 261, 513
\reference{} Smail I., Hogg D.W., Blandford R., Cohen J.G., Edge A.C., 
        Djorgovski S.G., 1995, MNRAS, 277, 1
\reference{} Smail I., Ivison R.J., Blain A.W., 1997, ApJ, 490, 5
\reference{} Smail I., Edge A.C., Ellis R.S., Blandford R.D.,
	1998, MNRAS, 293, 124
\reference{} Smail I., Ivison R.J., Kneib J.-P., Cowie L.L.,
	Blain A.W., Barger A.J., Owen F.N., Morrison G.,
	1999, MNRAS, 308, 1061
\reference{} Smail I., Ivison R.J., Blain A.W., Kneib J.-P.,
	2001, MNRAS, submitted
\reference{} Smith G.P., Smail I., Kneib J.-P., Czoske O., Ebeling H.,
	Edge A.C., Pell\'o R., Ivison R.J., Packham C., Le-Borgne J.-F., 2001,
	MNRAS, submitted (S01)
\reference{} Treu T., Stiavelli M., 1999, ApJ, 524, L27
\reference{} Veilleux S., Osterbrock D.E., 1987, ApJS, 63, 295
\end{references}
\end{document}